\documentclass[10pt,conference]{IEEEtran}
\IEEEoverridecommandlockouts
\usepackage{cite}
\usepackage{amsmath,amssymb,amsfonts}
\usepackage{algorithmic}
\usepackage{graphicx}
\usepackage{tabularx}
\usepackage{multirow}
\usepackage{subfigure}
\usepackage{hyperref}
\usepackage{xurl}
\usepackage{xargs}                      
\usepackage{verbatim}
\usepackage[dvipsnames]{xcolor}
\usepackage[colorinlistoftodos,prependcaption,textsize=small]{todonotes}
\newcommandx{\julia}[2][1=]{\todo[inline,linecolor=ForestGreen,backgroundcolor=ForestGreen!25,bordercolor=ForestGreen,#1]{Julia: #2}}
\usepackage{textcomp}
\def\BibTeX{{\rm B\kern-.05em{\sc i\kern-.025em b}\kern-.08em
    T\kern-.1667em\lower.7ex\hbox{E}\kern-.125emX}}

\usepackage[nolist,nohyperlinks]{acronym}
\newacro{5G}{Fifth Generation}
\newacro{5GC}{5G Core}
\newacro{5GS}{5G System}
\newacro{AMF}{Access and Mobility Management Function}
\newacro{BC}{Boundary Clock}
\newacro{BS}{Base Station}
\newacro{CF}{Correction Field}
\newacro{CNC}{Centralized Network Controller}
\newacro{CPE}{Customer Premise Equipment}
\newacro{DL}{Downlink}
\newacro{DS-TT}{Device-Side Time Sensitive Networking Translator}
\newacro{E2E}{End-to-End}
\newacro{FDD}{Frequency-Division Duplexing}
\newacro{FR}{Free-running}
\newacro{FRER}{Frame Replication and Elimination for Realibility}
\newacro{GM}{Grandmaster}
\acrodefplural{GMs}{Grandmasters}
\newacro{gNB}{Next-Generation Node B}
\newacro{GNSS}{Global Navigation Satellite System}
\newacro{gPTP}{generic Precision Time Protocol}
\newacro{IP}{Internet Protocol}
\newacro{IIoT}{Industrial Internet of Things}
\newacro{IoT}{Internet of Things}
\newacro{QoE}{Quality of Experience}
\newacro{QoS}{Quality of Service}
\newacro{L2}{Layer 2}
\newacro{L3}{Layer 3}
\newacro{NPN}{Non-Public Network}
\newacro{NR}{New Radio}
\newacro{NW-TT}{Network-Side Time Sensitive Networking Translator}
\newacro{P2P}{Peer-to-Peer}
\newacro{PCF}{Policy Control Function}
\newacro{PLC}{Programmable Logic Controller}
\acrodefplural{PLCs}{Programmable Logic Controllers}
\newacro{PPS}{Pulse Per Second}
\newacro{PTP}{Precision Time Protocol}
\newacro{RAN}{Radio Access Network}
\newacro{RRC}{Radio Resource Control}
\newacro{SCS}{Sub-Carrier Spacing}
\newacro{SFN}{System Frame Number}
\acrodefplural{SFN}{System Frame Numbers}
\newacro{SIB}{System Information Block}
\newacro{SMF}{Session Management Function}
\newacro{TAS}{Time-Aware Shaper}
\newacro{TC}{Transparent Clock}
\newacro{TDD}{Time Division Duplex}
\newacro{TSN}{Time Sensitive Networking}
\newacro{TSN AF}{TSN Application Function}
\newacro{TT}{TSN Translator}
\acrodefplural{TT}{TSN Translators}
\newacro{UDP}{User Datagram Protocol}
\newacro{UE}{User Equipment}
\newacro{UL}{Uplink}
\newacro{UPF}{User Plane Function}
\newacro{VLAN}{Virtual Local Area Network}
\acrodefplural{VLAN}{Virtual Local Area Networks}
\newacro{vPLC}{virtualized PLC}
\acrodefplural{vPLCs}{virtualized PLCs}
\newacro{VXLAN}{Virtual Extensible Local Area Network}

\begin{document}

\title{Empirical Evaluation of a 5G Transparent Clock for Time Synchronization in a TSN-5G Network\\
\thanks{This work has been financially supported in part by the Ministry for Digital Transformation and of Civil Service of the Spanish Government through the TSI-063000-2021-28 project, and by the European Union through the Recovery, Transformation and Resilience Plan - NextGenerationEU, and by grant PID2022-137329OB-C43 funded by MICIU/AEI/ 10.13039/501100011033 and by ERDF/EU,  by the Spanish Ministry of Universities (FPU Grant 21/04225) and by the Spanish Ministry of Universities (FPU Grant 20/02621).}
}

\author{\IEEEauthorblockN{Julia Caleya-Sanchez\IEEEauthorrefmark{1}\IEEEauthorrefmark{2}, Pablo Muñoz\IEEEauthorrefmark{1}\IEEEauthorrefmark{2}, Jorge Sánchez-Garrido\IEEEauthorrefmark{3}, Emilio Florentín\IEEEauthorrefmark{3}, Felix Delgado-Ferro\IEEEauthorrefmark{1}\IEEEauthorrefmark{2}, \\ Pablo Rodriguez-Martin\IEEEauthorrefmark{1}\IEEEauthorrefmark{2}
and Pablo Ameigeiras\IEEEauthorrefmark{1}\IEEEauthorrefmark{2}}    
\IEEEauthorblockA{\IEEEauthorrefmark{1}Department of Signal Theory, Telematics and Communications (DTSTC), University of Granada, Spain\\
\IEEEauthorrefmark{2}Research Center on Information and Communication Technologies, University of Granada, Spain.\\
\IEEEauthorrefmark{3}Safran Electronics and Defense Spain, Granada, Spain\\}
Emails: jcaleyas@ugr.es, pabloml@ugr.es, jorge.sanchez@nav-timing.safrangroup.com, \\emilio.florentin@nav-timing.safrangroup.com, felixdelgado@ugr.es, pablorodrimar@ugr.es, pameigeiras@ugr.es}

\maketitle

\begin{abstract}

Time synchronization is essential for industrial IoT and Industry 4.0/5.0 applications, but achieving high synchronization accuracy in Time-Sensitive Networking (TSN)-5G networks is challenging due to jitter and asymmetric delays. 3GPP TS 23.501 defines three 5G synchronization modes: time-aware system, boundary clock (BC), and transparent clock (TC), where TC offers a promising solution. However, to the best of our knowledge, there is no empirical evaluation of TC in a TSN-5G network. This paper empirically evaluates an 5G end-to-end TC in a TSN-5G network, implemented on commercial TSN switches with a single clock. For TC development, we compute the residence time in 5G and recover the clock domain at the slave node. We deploy a TSN-5G testbed with commercial equipment for synchronization evaluation by modifying the Precision Timing Protocol (PTP) message transmission rates. Experimental results show a peak-to-peak synchronization of 500 ns, meeting the industrial requirement of $\leq 1 \ \mu$s, with minimal synchronization offsets for specific PTP message transmission rates.

\end{abstract}


\section{Introduction}
\label{sec:Introduction}

Time synchronization distribution is an essential service for the \ac{IIoT}, playing a crucial role in the evolution towards Industry 4.0 and 5.0. Applications in these sectors depend on precise synchronization for reliable communication, low latency and operational efficiency to optimize automation and interaction between robots, sensors and machinery, as well as accurate sensor data analysis. 
The \ac{TSN} standard, 
defined by IEEE 802.1 \cite{IEEE802.1_Group}, enables synchronization accuracy below $100 \ \mu s$ \cite{Godor2020}, ensuring deterministic and reliable operation. 
However, the growing need for mobility in industrial, \ac{IoT} and robotics sectors presents new challenges, as \ac{TSN} was conceived for wired networks. As a solution, 3GPP Release 16 proposes integrating \ac{5G} with \ac{TSN} to provide wireless communication and mobile flexibility without compromising synchronization requirements. 

\ac{TSN} functionalities such as IEEE 802.1 Qbv, IEEE 802.1 Qcc and IEEE 802.1 Qci rely on accurate time synchronization. However, \ac{5G} introduces challenges that degrade synchronization accuracy, such as 
high levels of jitter, asymmetries in \ac{UL} and \ac{DL} delays, variable processing times in devices like the \ac{gNB} and \ac{UPF}, and retransmissions. 
TS 22.104 \cite{3gpp_ts22104} 
establishes the \ac{5GS} requirements for industrial and \ac{IIoT} networks, specifying a maximum time synchronization error of $900$ ns and support for up to $32$ simultaneous working clock domains. These domains provide high-precision time to the devices within them, allowing \ac{TSN} devices to manage multiple clock domains simultaneously. While a global clock domain provides the date and time to the devices belonging to that domain. In some cases, the working and global clock domains may overlap. Meanwhile, 3GPP TS 23.501 \cite{TS123501_V18} defines several configurations to enable \ac{TSN} synchronization in the \ac{5GS}
, which can be implemented as a time-aware system using the IEEE 802.1AS standard \cite{IEEE_8021_AS, SynchronizationIEEE802.1AS_Garner2011}, which uses the \ac{gPTP} protocol, or configured as a \ac{BC} or \ac{TC} following the IEEE 1588 standard \cite{StandadsIEEE1588_2019}. 

Several studies have addressed the challenges of time synchronization distribution in \ac{TSN}-\ac{5G} network. For example, Striffler et al. \cite{Striffler2021} analyze frequency drift and timing errors when \ac{5GS} operated as a \ac{TC}, which can lead to non-compliance with time synchronization requirements. While Wang et al. \cite{Wang2024} propose solutions to mitigate the effects of multi-\ac{gNB} competition, retransmissions, and mobility in a \ac{5GS} modeled as \ac{TC}. Shi et al. \cite{Shi2021} investigate synchronization errors in \ac{5GS} configured as \ac{BC}, focusing on the uncertainty due to reference time granularity and propagation delay estimation. Val et al. \cite{Val2022} demonstrate that, despite Wi-Fi variability, 
\ac{TSN}-compatible synchronization accuracies are achieved through hardware-level modifications, 
although this approach is currently infeasible for commercial \ac{5G} networks.


To our knowledge, there is no empirical evaluation of the \ac{TC} in an integrated \ac{TSN}-\ac{5G} network. Therefore, the objective of this paper is to empirically investigate the distribution of \ac{TSN} time synchronization in a \ac{TSN}-\ac{5G} network when the \ac{5GS} is configured as an \ac{E2E} \ac{TC}. Our main contributions are as follows:

\begin{list}{\textbullet}{
\setlength{\labelwidth}{1em}
     \setlength{\labelsep}{0.2em}
     \setlength{\itemsep}{0pt}
     \setlength{\leftmargin}{0.6em}
     \setlength{\rightmargin}{0cm}
     \setlength{\itemindent}{0cm}
}
    \item We analyze the various configurations of \ac{5GS} to support \ac{TSN} synchronization transport, considering its implementation in commercial single-clock equipment. 
    \item We implement the \ac{E2E} \ac{TC} on commercial \ac{TSN} switches. This implementation comprises two processes: the computation of the residence time  introduced by the \ac{5GS} and the recovery of the \ac{TSN} clock domain at the slave node. 
    \item We deploy a \ac{TSN}-\ac{5G} network testbed with commercial equipment to evaluate time synchronization performance. 
    \item We performed an experimental evaluation of the time synchronization accuracy on the implemented testbed. 
\end{list}

Our results indicate a peak-to-peak time synchronization of $500$ ns, meeting industrial requirements ($\leq 1 \ \mu s$). Also, we note that, at certain \ac{PTP} message transmission rates, time offsets can be observed that may affect the time synchronization but do not exceed the requirements. These results are independent of whether the synchronizing transmitter (master node) is synchronized to an external reference (\ac{GNSS}) or operates with its internal clock (\ac{FR}).

The rest of the paper is structured as follows: Section \ref{sec:Background} presents the background and an analysis of time synchronization options in \ac{TSN}-\ac{5G} networks. Section \ref{sec:SytemModel} describes the model and architecture of the integrated network. Section \ref{sec:TClock} details the design and operation of the \ac{TC}, along with \ac{PTP} messages analysis and offset calculation. Section \ref{sec:Testbed} describes the testbed and experimental setup. Section \ref{sec:ExperimentalResults} presents and discusses the results obtained. Finally, Section \ref{sec:Conclusion} concludes with the main contributions. 

\section{Background and Analysis of time synchronization options}
\label{sec:Background}

In the \ac{TSN}-\ac{5G} integrated network, there are two synchronization processes running in parallel, the \ac{5G} time synchronization process and the \ac{TSN} time synchronization process \cite{Godor2020, Integration5G_ACIA2021}, as shown in Figure \ref{fig:TSN-5G_Architecture}. The \ac{5G} synchronization process provides the temporal reference from the \ac{5G} \ac{GM} to the \ac{5GS} devices, such as the \ac{UPF}, \ac{gNB} and \ac{UE}. In paralell, the \ac{TSN} synchronization process provides the \ac{TSN} \ac{GM} temporal reference to the \ac{TSN} network devices. The two synchronization processes are considered to be independent, providing flexibility in the design and implementation of the synchronization process. Both processes are detailed below, but first we describe the \ac{5GS} architecture in the integrated \ac{TSN}-\ac{5G} network proposed by 3GPP. 

\subsection{Architecture of 5GS in the TSN-5G networks}
\label{subsec:TSN-5G_architecture}

\begin{figure}[tbp]
    \centering
    \includegraphics[width=\linewidth]{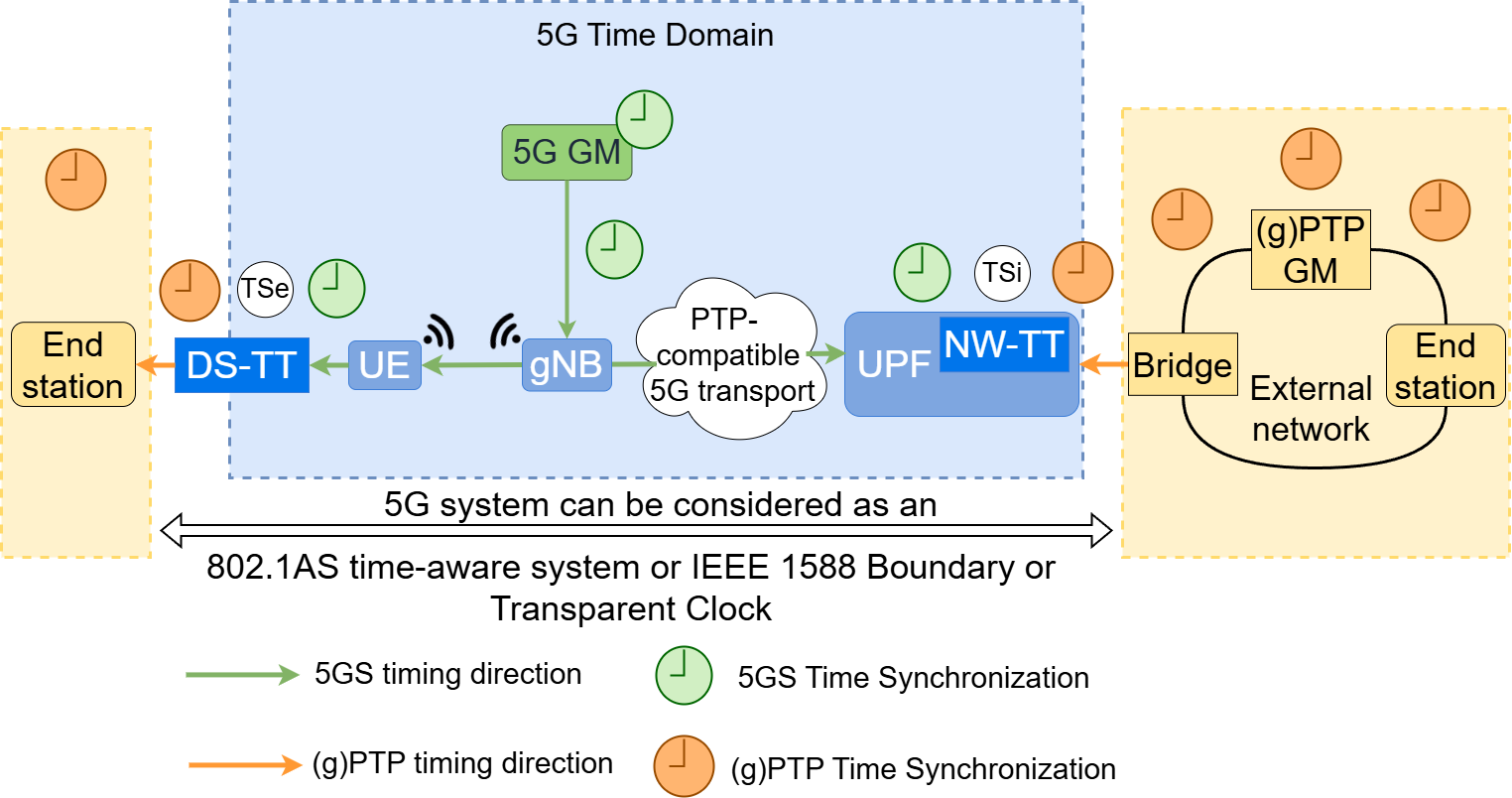}
    \caption{Architecture of the integrated TSN-5G network \cite{TS123501_V18}}
    \label{fig:TSN-5G_Architecture}
\end{figure}

According to 3GPP TS 23.501 \cite{TS123501_V18}, the \ac{5GS} acts as a \ac{PTP} instance in the \ac{TSN} network. As shown in Figure \ref{fig:TSN-5G_Architecture}, the external network consists of end stations and \ac{TSN} bridges, whose time reference is provided by a \ac{PTP} or \ac{gPTP} \ac{GM}. In the \ac{5G} domain, a \ac{5G} \ac{GM} synchronizes the \ac{gNB}, which distributes the time reference to the \ac{UE} over the radio link and to the \ac{UPF} via the \ac{PTP}-compliant \ac{5G} transport network. To integrate \ac{5G} with \ac{TSN}, the \ac{UPF} implements a \ac{TT} (\ac{NW-TT}) that interfaces with the external \ac{TSN} bridge, while the \ac{UE} connects to another \ac{TT} (\ac{DS-TT}) and this connects to a \ac{TSN} end station. These \ac{TT} ensure interoperability, allowing the \ac{5GS} to operate as a logical bridge in various modes without direct alignment with the \ac{TSN} \ac{GM}.


\subsection{The 5G synchronization process}
\label{subsec:Sync5G}

The \ac{5G} time synchronization process is based on the internal distribution of the \ac{5G} \ac{GM} reference across the \ac{5GS} transport network and the \ac{5GS} \ac{RAN} \cite{Patel2021}. 
A straightforward solution in the \ac{5G} transport network is to install \ac{GNSS} receivers in each \ac{gNB}, which offers $\pm 100$ ns accuracy \cite{Patel2021}, meeting the \ac{5GS} timing requirements. However, this method comes with high deployment and maintenance costs, indoor installation difficulties and vulnerability to jamming. Alternatively, synchronization protocols over packet networks, such as IEEE 1588 \cite{StandadsIEEE1588_2019}, with adapted profiles defined by ITU-T (e.g., G.8275.1 (\ac{PTP}-aware) and G.8275.2 (non-\ac{PTP}-aware)). 

In the \ac{5G} \ac{RAN}, synchronization between the \ac{5G} \ac{GM} and the \ac{UE} is achieved through the \ac{gNB}, as defined in TS 38.331. 
The \ac{gNB} continuously updates the \ac{5G} \ac{GM} reference \cite{Godor2020, Patel2021, TAPv2} and periodically transmits it to the \ac{UE} via \ac{SIB} or unicast \ac{RRC} messages, identified via a \ac{SFN}. The \ac{SIB}9 message, which contains the time information in GPS and UTC formats, is transmitted at the boundary between two \acp{SFN}, enabling synchronization of the devices. In addition, the \ac{5G} reference must be adjusted according to the cell size to compensate for \ac{DL} propagation delay and minimize reception uncertainty. Accurate delay calculation and obtaining the \ac{SIB}9 reference are essential aspects, although they are not addressed in this study. For more information, see \cite{TAPv2, Patel2021}. 

\subsection{The TSN synchronization process}
\label{subsec:SyncTSN}

TS 23.501 \cite{TS123501_V18} defines several modes in which the \ac{5GS} can be configured to operate as one \ac{PTP} instance \cite{Godor2020}, enabling \ac{TSN} time synchronization in an integrated \ac{TSN}-\ac{5G} network. Specifically, these clock modes are: 

\begin{list}{\textbullet}{
\setlength{\labelwidth}{1em}
     \setlength{\labelsep}{0.2em}
     \setlength{\itemsep}{0pt}
     \setlength{\leftmargin}{0.6em}
     \setlength{\rightmargin}{0cm}
     \setlength{\itemindent}{0cm}
}    
    \item \textbf{\ac{5GS} as a Time-Aware System:}
    The \ac{5GS} behaves as an IEEE 802.1AS-compliant node \cite{IEEE_8021_AS}, participating in time synchronization using the \ac{gPTP} protocol. \ac{NW-TT} and \ac{DS-TT} synchronize with both \ac{GM} clocks (\ac{TSN} \ac{GM} and \ac{5G} \ac{GM}), thus requiring two clocks, and manage \ac{gPTP} messages (transmitted in \ac{L2}) to ensure synchronization \cite{Godor2020}. 
    
    \item \textbf{\ac{5GS} as \ac{BC}:}
    The \ac{5GS} is configured as a \ac{BC} according to the IEEE 1588 standard \cite{StandadsIEEE1588_2019}, actively participating in synchronization. That is, the \ac{5GS} maintains the \ac{PTP} domain timescale and synchronizes the connected \ac{PTP} time receivers (e.g., \ac{DS-TT}, \ac{TSN} Slave, etc.), acting as the time source.
    This mode provides higher accuracy and robustness, allowing network scalability and lower impact of jitter and propagation delay in multi-device scenarios, but is more complex.
    
    \item \textbf{\ac{5GS} as \ac{TC}:}
    The \ac{5GS} behaves as a \ac{TC} according to the IEEE 1588 standard \cite{StandadsIEEE1588_2019}, without actively participating in time synchronization, but measures the residence time of \ac{PTP} messages within the network and uses it to correct them before forwarding them. The \acp{TT} manage the \ac{PTP} messages and calculate the delay within the \ac{5GS}.
    

    There are two types of \ac{TC}: \ac{P2P} or \ac{E2E}. In \ac{P2P} \ac{TC} \cite{IEEE802.1AS_NetworkIEEE1588_Garner2010} the \ac{5GS} measures the delay between two directly connected neighboring node using Peer Delay Messages 
    and calculates the residence time of the \ac{PTP} messages within the \ac{5G} network. The sum of the residence time and the link delay along the path is reported to the \ac{PTP} time receiver \cite{SynchronizationIEEE802.1AS_Garner2011}.
    The \ac{E2E} \ac{TC} calculates the total \ac{E2E} delay between the \ac{PTP} transmitter and receiver by exchanging \ac{PTP} messages. In the \ac{5GS} the residence time is calculated and the sum of the residence time over the entire path is added in the correction field of the \ac{PTP} messages. This sum of residence times is relayed to the \ac{PTP} time receiver. This way, the \ac{PTP} receiver can calculate the total compensation based on the sum of residence times in the correction field of the messages. 
    
    Therefore, \ac{P2P} \ac{TC} compensates for the latency between neighboring nodes while \ac{E2E} \ac{TC} compensates for the latency of the entire \ac{E2E} path. \ac{P2P} \ac{TC} requires IEEE 1588 compatibility on all network devices, \ac{E2E} \ac{TC} does not. 

\end{list}

In this paper, we opt for the \ac{E2E} \ac{TC} mode in \ac{5GS}, as it allows transmission of multiple working clock domains from different \ac{GM}s, a key requirement in industrial environments. The IEEE 802.1AS standard, requiring \ac{L2} transmission, has limitations for integration into commercial equipment. However, it has been demonstrated that an IEEE 802.1AS network can operate as distributed clock (\ac{BC}, ordinary or \ac{TC}) according to IEEE 1588 \cite{IEEE802.1AS_NetworkIEEE1588_Garner2010}, facilitating the transport of multiple clock signals between domains. Moreover, the \ac{TC} by operating independently, without the need to synchronize its own clocks simplifies the integration of \ac{5GS} into existing \ac{TSN} networks or enables the incorporation of new \ac{5G} operators, thus ensuring greater flexibility in the deployment of \ac{TSN}-\ac{5G} network.


\section{System Model}
\label{sec:SytemModel}

\begin{figure}[tbp]
    \centering
    \includegraphics[width=\linewidth]{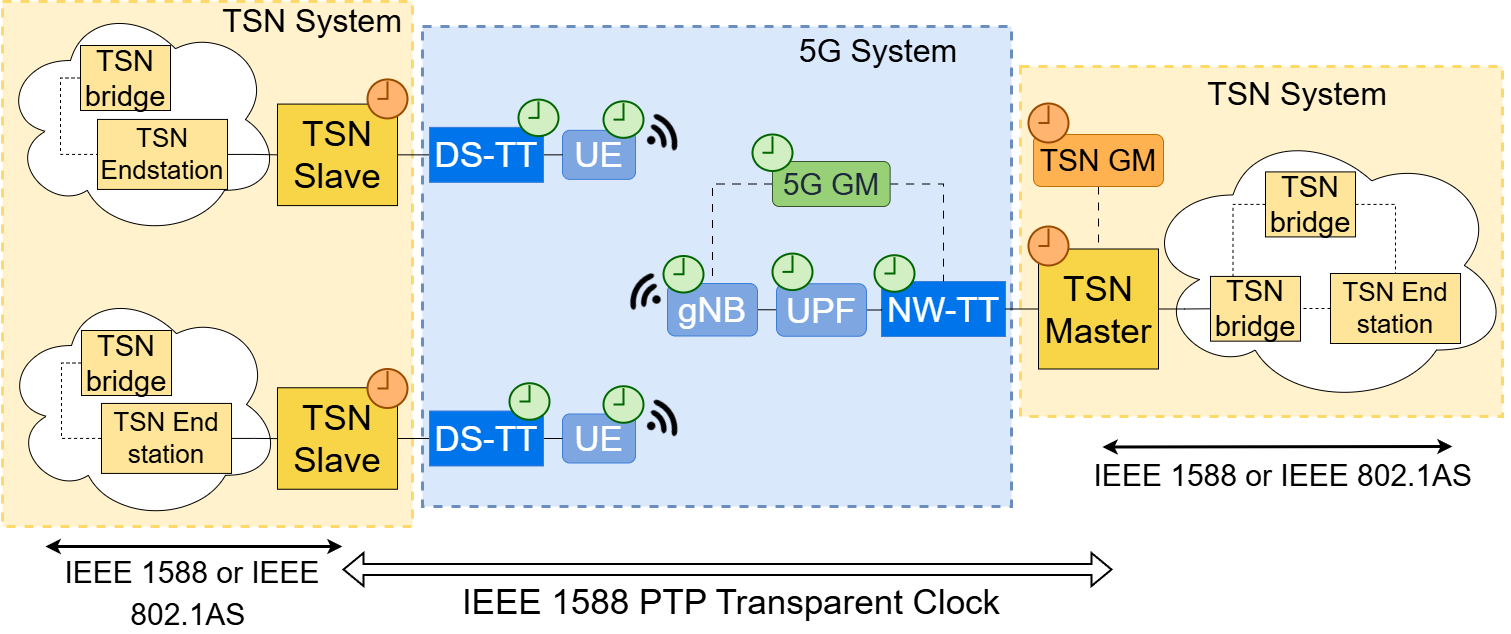}
    \caption{TSN-5G system model for time synchronization}
    \label{fig:SysteModel_Arq}
\end{figure}

We consider the \ac{TSN}-\ac{5G} system model as illustrated in Figure \ref{fig:SysteModel_Arq}. The \ac{TSN} system includes a \ac{TSN} Master, synchronized with the \ac{TSN} \ac{GM}, which distributes the time reference to other \ac{TSN} bridges and/or \ac{TSN} end station. This synchronization is extended to a \ac{TSN} Slave in another \ac{TSN} domain, interconnected through the \ac{5G} network. This \ac{TSN} Slave, once synchronized, distributes the synchronization to other \ac{TSN} bridges or \ac{TSN} end stations.
The \ac{5GS} incorporates a \ac{5G} \ac{GM}, which directly provides the time reference to the \ac{gNB} and \ac{NW-TT}, both connected through the \ac{UPF}. Since commercial equipment does not integrate a \ac{UPF} with \ac{NW-TT}, 
the \ac{NW-TT} is synchronized with the \ac{5G} \ac{GM} and connected directly to the \ac{TSN} Master. While the \ac{DS-TT}, also synchronized to the \ac{5G} \ac{GM} via the \ac{UE}, connects to the \ac{TSN} Slave. This differs from the model in Figure \ref{fig:TSN-5G_Architecture}, as in this case the \ac{NW-TT} and \ac{UPF} are separate devices, and both the \ac{NW-TT} and \ac{DS-TT} are directly connected to the \ac{TSN} Master and \ac{TSN} Slave, respectively.

Time synchronization in the \ac{TSN}-\ac{5G} network is based on IEEE 1588 compliant PTP frame transmission \cite{StandadsIEEE1588_2019}. In this paper, we use the \ac{E2E} \ac{TC} clock mechanism for time synchronization. We define $\mathcal{F}$ as the set of \ac{PTP} flows traversing the  \ac{TSN}-\ac{5G} network. Each \ac{PTP} flow $f_i \ \forall \ i \in \mathcal{F}$ represent a bidirectional communication between a \ac{PTP} transmitter node, synchronized with a \ac{GM} and a \ac{PTP} receiver node, not synchronized with the \ac{GM}. We refer as a \ac{PTP} transmitter node to the \ac{TSN} Master and as a \ac{PTP} receiver node to the \ac{TSN} Slave. These \ac{PTP} flows ensure that the \ac{PTP} transmitter and the \ac{PTP} receiver maintain the same time reference as the \ac{GM}. Therefore, each flow $f_i \ \forall \ i \in \mathcal{F}$ is transmitted from the \ac{TSN} Master to the \ac{TSN} Slaves, traversing the \ac{5G} network. In our case, we focus on a single \ac{PTP} flow $f_i$ to evaluate synchronization. This $f_i$ transmits a series of $p^{f_i}_{\mathcal{E}}$ \ac{PTP} packets between the \ac{TSN} Master and the Slave using the \ac{E2E} \ac{TC} mechanism detailed in Section \ref{sec:TClock}. For clarity, henceforth, the superindex $f_i$ will be omitted, meaning $p^{f_i}_{\mathcal{E}} = p_{\mathcal{E}}$ in all variables, since only a single \ac{PTP} flow is considered.
 
In this scenario, as previously explained, two synchronization processes coexist \cite{Integration5G_ACIA2021}: \ac{TSN}  synchronization and \ac{5G} synchronization.  
We consider that each device within the \ac{TSN}-\ac{5G} network has a single clock. The clocks of the \ac{TSN} Master and Slave are synchronized with the \ac{TSN} \ac{GM} reference, while the \acp{TT} are solely synchronized with the \ac{5G} \ac{GM}. In \ac{5G} synchronization, the \ac{5G} \ac{GM} reference is transmitted through the \ac{gNB} to the \ac{UE} and the \ac{UE} transmits it to the \ac{DS-TT}. In this paper, we do not focus on the \ac{5G} clock recovery by the \ac{UE} and its subsequent transmission to the \ac{DS-TT}, therefore, we assume perfect recovery from \ac{SIB}9 for synchronization at the \ac{UE} and subsequently at the \ac{DS-TT}.

In \ac{TSN} synchronization, the $p_{\mathcal{E}}$ include a timestamp. Specifically, each $p_e \ e \in \mathcal{E}$ contains a timestamp that records the moment the \ac{TSN} Master sends the packet $e$. However, due to delays caused by software and hardware packet processing in the \ac{TSN} Master, the timestamp in $p_e$ may lack precision. Then, the transmission timestamp of $p_e$ is included in $p_{e+1}$, recording the precise moment ($t_1$) when $p_e$ exits the \ac{TSN} Master. When the \ac{TSN} Slave receives $p_e$, it records the timestamp ($t_2$).

In addition, $p_{\mathcal{E}}$ packets when traversing the \ac{5G} network experience variable delays (jitter), compromising the timing accuracy and affecting the correct operation of the \ac{TSN} nodes. This present a challenge to accurately estimate the time reference. To address this issue, the delay experienced by the $p_{\mathcal{E}}$ while traversing the \ac{5G} network is calculated. This delay is referred to as the residence time in the \ac{5G} network ($d_{res}$). Its calculation involves generating timestamps for certain $p_e$ packet at the \acp{TT}. The \ac{NW-TT} generates an ingress timestamp ($t_{in}$) based on the \ac{5GS} reference time, while the \ac{DS-TT} generates an egress timestamp ($t_{eg}$) also based on the \ac{5GS} reference time. The residence time ($d_{res}$) is defined as:
\begin{equation}\label{eq:Cal_tiempo_resid_general}
    d_{res} = t_{eg} - t_{in}
\end{equation}
Since the \ac{5G} network is asymmetric in its links, meaning the \ac{UL} and \ac{DL} delays differ depending on the radio channel's characteristics, a specific residence time is defined for each link. The residence time for the \ac{UL} is denoted as $d_{res,up}$, and for the \ac{DL} as $d_{res,down}$.

Each \ac{TSN} node incorporates an internal oscillator that may introduce noise to the device's internal clock due to physical disturbances, causing clock skew. To mitigate this clock drift, \ac{PTP} messages are generated with cycle $T$.

\section{Transparent Clock Design}
\label{sec:TClock}

We focus on the distribution of time synchronization in an integrated \ac{TSN}-\ac{5G} network, implementing \ac{E2E} \ac{TC} mode in the \ac{5GS}. With this approach, \ac{5G} devices do not need to synchronize with the \ac{TSN} \ac{GM} time reference, as the \ac{5G} common time reference is used to accurately calculate the residence times ($d_{res,up}$, $d_{res,down}$) of $p_{\mathcal{E}}$ traversing the network. 
As explained in Section \ref{sec:SytemModel}, it is essential that the \ac{NW-TT} and \ac{DS-TT} share the same time reference derived from the \ac{5G} \ac{GM}. This allows that the \ac{TC} mechanism to operate in a manner equivalent to a standalone switch, ensuring consistency in the calculation of residence times from the 
timestamps generated on devices located at opposite ends of \ac{5GS}. Any failure to propagate a common time reference could lead to inconsistencies or invalid synchronization results between \ac{TSN} systems at the edges. The following section details the \ac{PTP} messages exchanged and the procedure to be followed to calculate the offset and residence time.

\subsection{Messages, procedures and offset calculation}
\label{subsec:Procedure}

\begin{figure}[tbp]
    \centering
    \includegraphics[width=0.91\linewidth]{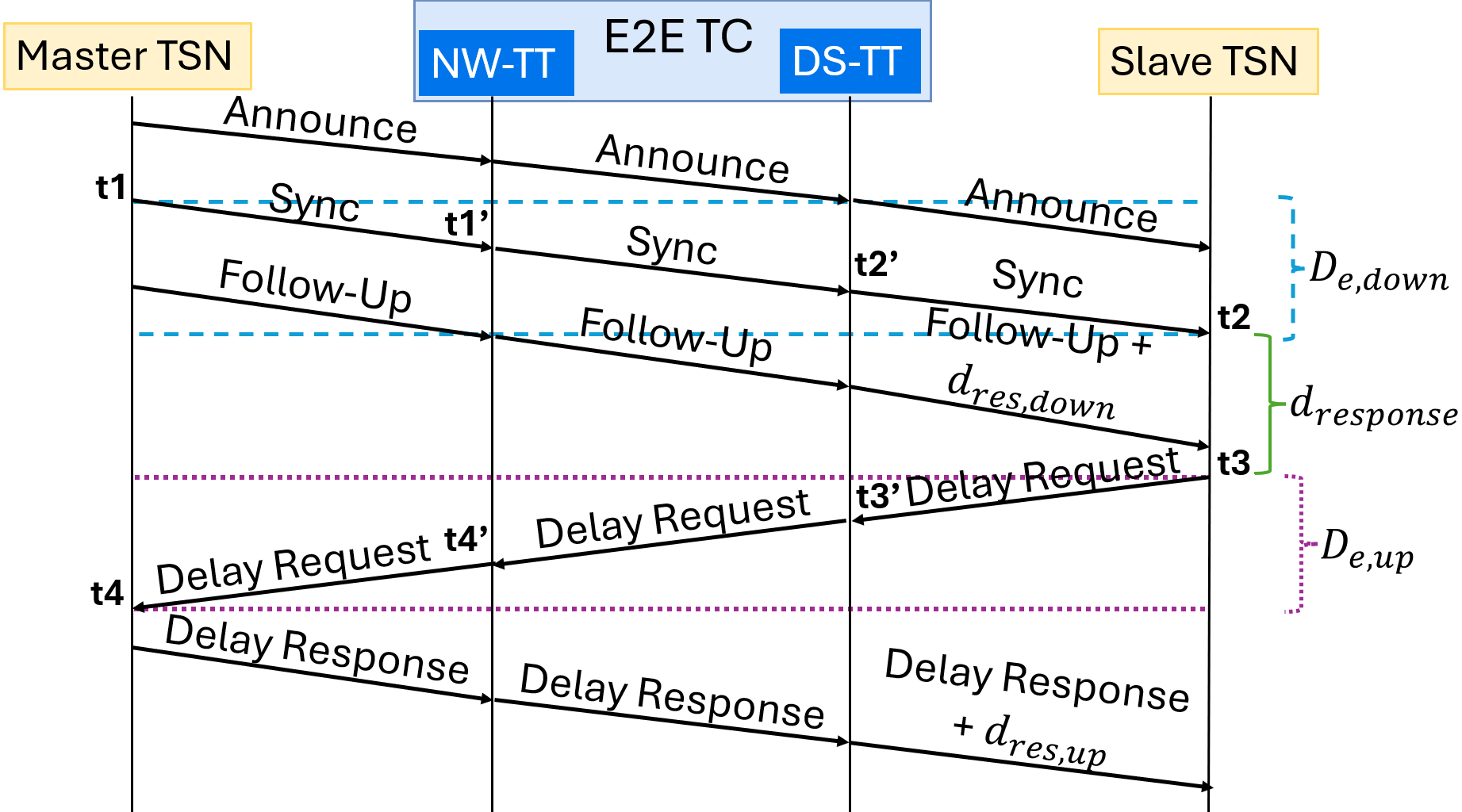}
    \caption{End-to-End Transparent Clock messages exchange}
    \label{fig:MessagesE2ETC}
\end{figure}

The \ac{E2E} \ac{TC} solution implemented in the \ac{5GS} exchanges \ac{PTP} messages between the nodes of the distributed architecture, as illustrated in Figure \ref{fig:MessagesE2ETC}.  The \ac{TSN} Master initiates synchronization by transmitting ``Announce'' messages, followed by ``Sync'' and ``Follow-Up''. The \ac{TSN} Slave then sends ``Delay Request'' and receives a ``Delay Response'' in reply. This exchange enables the measurement of the residence time on the \ac{5GS}, through the time stamping and delay estimation capabilities of the \ac{NW-TT} and \ac{DS-TT}, and the recovery of the \ac{TSN} clock domain at the \ac{TSN} Slave. Both functionalities are key to the operation of the \ac{TC} and to perform \ac{TC} on commercial switches, we have implemented both functionalities in these devices. Each functionality is detailed below: 

\renewcommand{\labelenumi}{{\theenumi})}
\newcounter{boxlblcounter}  
\newcommand{\makeboxlabel}[1]{{#1)}}
\newenvironment{boxlabel}
  {\begin{list}
    {\arabic{boxlblcounter}}
    {\usecounter{boxlblcounter}
     \setlength{\labelwidth}{1em}
     \setlength{\labelsep}{0.2em}
     \setlength{\itemsep}{0pt}
     \setlength{\leftmargin}{0.6em}
     \setlength{\rightmargin}{0cm}
     \setlength{\itemindent}{0cm} 
     \let\makelabel=\makeboxlabel
    }
  }
{\end{list}}

\begin{boxlabel}
    \item \underline{Calculation of the \ac{5GS} residence time}: \ac{NW-TT} and \ac{DS-TT} assign timestamps to the ``Sync'' packets at the ingress ($t_{1'}$) and egress ($t_{2'}$) of the \ac{5GS}, respectively. These timestamps determine the \ac{DL} residence time ($d_{res,down}$), as detailed in equation \eqref{eq:TiempoRetardoTC}. The value of $d_{res,down}$ is temporarily stored in \ac{DS-TT} until the arrival of the ``Follow-Up'' packet, where \ac{DS-TT} updates the \ac{CF} with the value of $d_{res,down}$. This procedure allows the \ac{TSN} Slave to compensate for the delay introduced by the ``Sync'' packets within the \ac{5GS} for its estimates of the total link delay. An analogous process is applied for the ``Delay Request'' and ``Delay Reply'' packets, allowing the \ac{UL} residence time ($d_{res,up}$) estimation using the timestamps of ``Delay Request'' and the \ac{CF} update in ``Delay Reply'', as shown in Figure \ref{fig:MessagesE2ETC}. The \ac{UL} and \ac{DL} residence time are calculated as:
    \begin{equation}\label{eq:TiempoRetardoTC}
        \begin{split}
            d_{res,down} = t_{2'} - t_{1'} \\
            d_{res,up} = t_{4'} - t_{3'} 
        \end{split}
    \end{equation}

    \item \underline{Clock recovery at the \ac{TSN} Slave}: several operations are necessary \cite{StandadsIEEE1588_2019}. 
    First, we decouple the delay of the \ac{PTP} packets sent from the \ac{TSN} Master to the \ac{TSN} Slave from the variable residence time of the \ac{5GS}. That is, due to the residence time calculation, the \ac{5GS} introduces variable jitter in the \ac{UL} and \ac{DL}, generating different delays for \ac{PTP} packets. 
    We call $D_{e,down}$ the \ac{DL} delay of the ``Sync'' packet, while $D_{e,up}$ is the \ac{UL} delay of the ``Delay Request'' packet. The sum of both delays determines the $p_{\mathcal{E}}$ transmission delay ($D_{\mathcal{E}}$). $D_{e,down}$, $D_{e,up}$ and $D_{\mathcal{E}}$ are calculated as: 
    \begin{gather}\label{eq:RetardoUpDown}
        D_{e,down} = t_{2} - t_{2'} + t_{1'} - t_{1} = t_2 - t_1 - d_{res,down} \\
        D_{e,up} = t_{4} - t_{4'} + t_{3'} - t_{3} = t_4 - t_3 - d_{res,up} \\
        D_{\mathcal{E}} = D_{e,down} + D_{e,up}
    \end{gather}
    $d_{res,down}$ and $d_{res,up}$ mitigate the associated problem of having different latencies in \ac{UL} and \ac{DL}. The \ac{TSN} Slave requires a response time ($d_{response}$) after receiving the ``Sync'' before generating the ``Delay Request'', as shown in Figure \ref{fig:MessagesE2ETC}. This $d_{response}$ is calculated as:
    \begin{equation}\label{eq:Trespuesta}
        \begin{split}
             d_{response} = t_{3} - t_{2}
        \end{split}
    \end{equation}

    The \ac{TSN} Slave calculates the time offset ($\beta$) with respect to the \ac{TSN} Master from all timestamps, the residence time and the response time, as follows: 
    \begin{equation}
        \begin{split}
            \beta = (t_{2} - t_{1}) - \frac{D_{\mathcal{E}}}{2} - d_{res,down} 
        \end{split}
    \end{equation}
\end{boxlabel}

Thus, the \ac{TSN} Slave compensates for its clock offset to set the \ac{TSN} \ac{GM} time. This procedure compensates for latency variability in the propagation of time references in \ac{5GS}, ensuring accurate synchronization throughout the network, essential for industrial applications and the \ac{IIoT}. Both functionalities have been implemented in commercial equipment that make up the \ac{TT} and the \ac{TSN} Slave.

\section{Testbed and Experimental Setup}
\label{sec:Testbed}

\begin{figure}[tpb]
    \centering
    \includegraphics[width=\columnwidth]{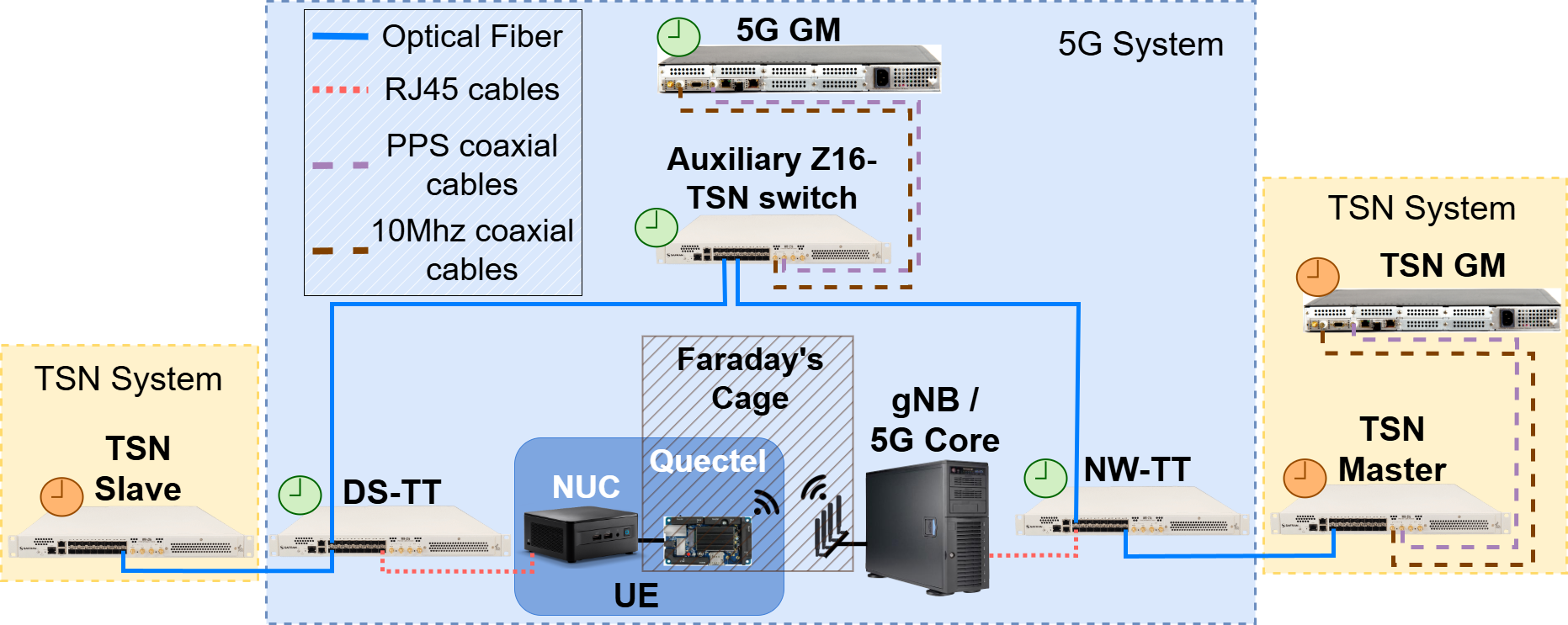}
    \caption{TSN-5G testbed}
    \label{fig:Experimental_Setup}
\end{figure}

We implemented a \ac{TSN}-\ac{5G} testbed to evaluate time synchronization performance, illustrated in Figure \ref{fig:Experimental_Setup}, where we only considered \ac{PTP} traffic. The \ac{TSN} systems are composed of \ac{TSN} Master and \ac{TSN} Slave nodes, implemented on commercial \ac{TSN} switches (Z16 from Safran). These devices integrate time-sensitive networking features according to the specifications set forth in \cite{IEEE802.1_Group}. \ac{TSN} Slave contains modifications to implement the \ac{TSN} clock recovery functionality, as detailed in Section \ref{subsec:Procedure}. For \ac{PTP} packet encapsulation and exchange, these devices use the \ac{UDP} protocol over IPv4 in unicast mode with the \ac{E2E} mechanism, a modality widely adopted in \ac{5G} networks with commercial devices. 
In this UDP/IPv4 mode, the source and destination IP addresses of the \ac{PTP} packets are the addresses assigned to the \ac{TSN} Master and \ac{TSN} Slave ports, respectively. The \ac{PTP} packet transmission rate varies depending on the experiment. The \ac{TSN} Master is synchronized with the \ac{TSN} \ac{GM}, implemented by a time and frequency reference server (Safran's Secure Sync). The time synchronization between the \ac{TSN} GM and the \ac{TSN} Master is performed by coaxial cables, using the \ac{PPS} and $10$ MHz signals. 

\ac{5GS} consists of the \acp{TT}, also implemented with commercial \ac{TSN} switches (Safran's Z16), both modified to implement \ac{TC}. These \acp{TT} are synchronized with the \ac{5G} clock provided by a second Safran Secure Sync server, which acts as the \ac{5G} \ac{GM}. Since the \ac{5G} \ac{GM} only has a single \ac{PPS} and $10$ MHz output, an auxiliary \ac{TSN} switch (Z16-TSN) is required to distribute the time reference to the \acp{TT}. This switch is synchronized with the \ac{5G} \ac{GM} via coaxial cables connected to the \ac{PPS} and $10$ MHz signals, see Figure \ref{fig:Experimental_Setup}. Subsequently, the switch transmits the time reference to the \ac{NW-TT} and \ac{DS-TT} via \ac{L2}, using an \ac{E2E} mechanism with a \ac{PTP} message transmission rate of 1 packet/s. The \ac{5GS} also consist of a \ac{BS} and a \ac{5G} core, both integrated in a PC with two $50$ MHz PCLe SDR Amarisoft cards and an AMARI NW 600 license. The \ac{BS} operates in the frequency band \textit{n78}, using a \ac{SCS} of 30 kHz and a \ac{TDD} mode, with a \ac{TDD} pattern of 1,0. The \ac{UE} consists of a Quectel RM500Q-GL modem connected to an Intel NUC 10 NUC10i7FNKN, which acts as a \ac{CPE}. This device contains an Intel i7-10710U processor with 16 GB of RAM and 512 GB of SSD memory, and runs Ubuntu 22.04.

All experiments are performed inside a LABIFIX Faraday cage, where the \ac{BS} antennas are connected to a SDR via SMA connectors and the Quectel modem is connected via USB. This cage avoids radiation within the licensed frequency bands. The connections between the rest of the equipment are made using 1 Gbps optical fiber, except for the connections between the \ac{NW-TT} and \ac{DS-TT} to the \ac{gNB} and \ac{UE}, respectively, which use 1 Gbps RJ-45 cables, as shown in Figure \ref{fig:Experimental_Setup}. 

\section{Experimental Results}
\label{sec:ExperimentalResults}

We have performed two experiments to evaluate the performance of \ac{TC} and the accuracy of \ac{TSN} time synchronization. The first experiment analyzes the synchronization between the \ac{TSN} Master and the \ac{TSN} Slave when varying the \ac{PTP} packet transmission rates. The second experiment compares the time synchronization when the \ac{TSN} Master is synchronized to the \ac{TSN} \ac{GM}, locked to a \ac{GNSS} reference, versus when operating in \ac{FR}, using its own internal clock. In both cases, we measured the offset between the \ac{TSN} Master and \ac{TSN} Slave clocks for $20$ min with a high precision counter. Table \ref{tab:tc_ptp_settings} summarizes the configurations used in the experiments.


\begin{table}[tbp]
\caption{Overview of $p_{\mathcal{E}}$ rate settings per TC experiment}
\label{tab:tc_ptp_settings}
\begin{tabular}{|c|m{1.1cm}|m{0.7cm}|m{0.8cm}|m{0.8cm}|m{0.8cm}|m{0.7cm}|}
\hline
\textbf{Test ID} & \textbf{Announce} & \textbf{Sync} & \textbf{Follow-Up} & \textbf{Del. Req.} & \textbf{Del. Resp.} & \textbf{Sync. Mode} \\ \hline
    \textit{0} & 1 p/2s & 1 p/2s & 1 p/2s & 1 p/2s & 1 p/2s & \\ \cline{1-6}
    \textit{1} & 1 p/s & 1 p/s & 1 p/s & 1 p/s & 1 p/s & \\ \cline{1-6}
    \textit{2} & 2 p/s & 2 p/s & 2 p/s & 2 p/s & 2 p/s & \multirow{-3}{*}{GNSS}     \\ \hline
    \textit{3} & 2 p/s & 2 p/s & 2 p/s & 2 p/s & 2 p/s & FR \\ \hline
\end{tabular}
\end{table}

\subsection{Experiment 1: Impact of PTP message transmission rate}
\label{subsec:Experiment1}

\begin{figure}
    \centering
    \includegraphics[width=\columnwidth]{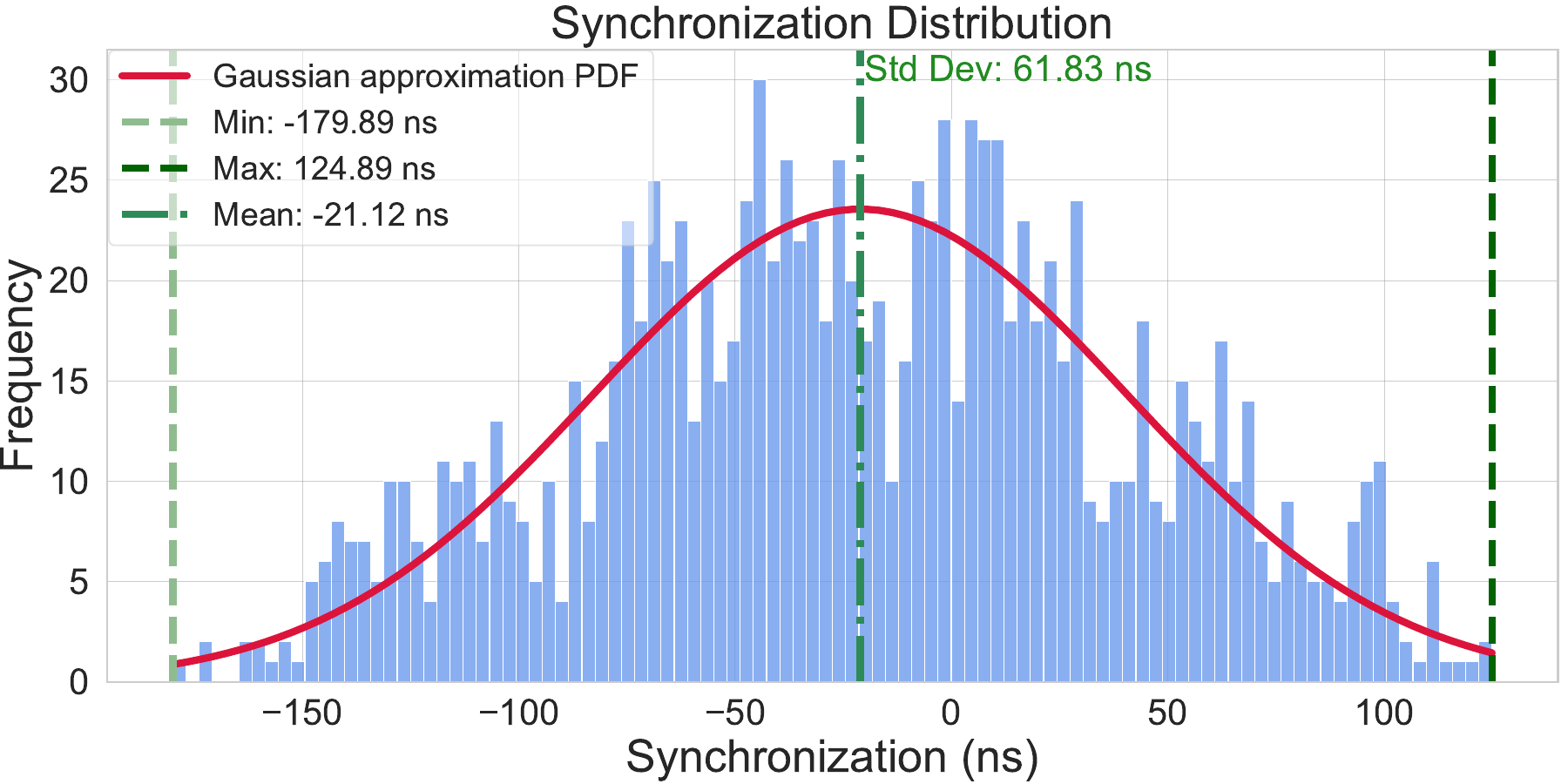}
    \caption{Synchronization distribution for 1 packet/s rate}
    \label{fig:Experimental1_1ps}
\end{figure}

The measured offset results of the \ac{TC} mechanism, with the \ac{GM} synchronized to a \ac{GNSS}, are presented in Table \ref{tab:tc_ptp_results_gnss_fr} for the tests \textit{``0''} to \textit{``2''}, showing the average, maximum, minimum and standard deviation of the offset in each case. 
In particular, the measured offset at a rate of $1$ p/s is illustrated in Figure \ref{fig:Experimental1_1ps}, whose distribution follows approximately a Gaussian distribution. Similar behavior is observed for the rest of the tests. 
The data demonstrate very good synchronization performance for the \ac{E2E} \ac{TC} mechanism, with overall jitter below $500$ ns and offset in the range of hundreds of nanoseconds. The best performance is obtained in the test \textit{ ``1''}, while lower transmission rates generate slower offset correction, degrading accuracy. Conversely, higher rates cause packet accumulation in the \acp{TT}, which alters the residence time measurement at \ac{5GS}, inducing errors in the \ac{TSN} Slave clock correction.
Remarkably, all test meets the synchronization requirements of industrial applications, validating the effectiveness of our the distributed \ac{E2E} \ac{TC} system, even in a high variability and non-deterministic medium, such as \ac{5GS}.


\begin{table}[tbp]
\begin{center}
    \caption{Overview of TC synchronization performance for $p_{\mathcal{E}}$ settings}
    \label{tab:tc_ptp_results_gnss_fr}
    \begin{tabular}{|c|c|c|c|>{\centering\arraybackslash}m{4.305em}|c|}
    \hline
    \textbf{Test ID} & \textbf{Mean (ns)} & \textbf{Max (ns)} & \textbf{Min (ns)} & \textbf{Std. Dev. (ns)} & \textbf{Mode} \\ \hline
    \textit{0} & -18.70 & 269.92 & -279.80 & 108.58 & \multirow{3.5}{*}{GNSS} \\ \cline{1-5}
    \textit{1} & -21.13 & 124.89 & -179.89 & 61.83 &         \\ \cline{1-5}
    \textit{2} & 238.28 & 437.09 & 29.80 & 80.80 &                       \\ \hline
    \textit{3} & -70.01 & 167.15 & -268.26 & 76.78 & FR \\ \hline
    \end{tabular}
\end{center}
\end{table}

\subsection{Experiment 2: Performance comparison of TSN GM}
\label{subsec:Experiment2}

The measured offset results when \ac{TSN} Master operates in \ac{FR} are presented in test \textit{``3''} of Table \ref{tab:tc_ptp_results_gnss_fr}, showing a Gaussian-like distribution, similar to that illustrated in Figure \ref{fig:Experimental1_1ps}.
Comparison of results with test \textit{``2''} shows an equivalent temporal accuracy. Although \ac{FR} seems to offer slightly better stability, this could be attributed to the absence of adjustments to the internal oscillator of the \ac{TSN} Master to follow the \ac{GNSS} reference, which avoids additional fluctuations. However, this does not imply higher accuracy as it is not compared to the global time reference.
Since the \ac{5GS} introduces significant latency variations, the differences in offset between \ac{FR} and \ac{GNSS} are negligible. Thus, we conclude that the \ac{E2E} \ac{TC} mechanism operates equivalently under both references without significant impacts on synchronization.

\section{Conclusion}
\label{sec:Conclusion}
This paper empirically evaluates an \ac{E2E} \ac{TC} in a \ac{TSN}-\ac{5G} network, implemented on commercial \ac{TSN} switches with a single clock. The solution contains the computation of the residence time within \ac{5GS} (\ac{NW-TT} and \ac{DS-TT}), and the recovery of the \ac{TSN} clock domain at the slave node. We have deployed a \ac{TSN}-\ac{5G} testbed with commercial equipment to analyze time synchronization at different \ac{PTP} message rates. The results show a peak-to-peak accuracy of $500$ ns, meeting industrial requirements, and show that certain transmission rates can induce offsets without exceeding the allowed margins, regardless of the reference (\ac{GNSS} or \ac{FR}).
This work represents a first step to demonstrate the feasibility of \ac{E2E} \ac{TC} in an integrated \ac{TSN}-\ac{5G} network. Future research will explore synchronization under traffic load and extraction of the \ac{SIB}9 reference to align the \ac{UE} with the \ac{5G} \ac{GM} clock of \ac{gNB}. 

\bibliographystyle{IEEEtran}
\bibliography{bibliografia}


\end{document}